\documentstyle[11pt,newpasp,twoside,epsf]{article}
\def\edcomment#1{\iffalse\marginpar{\raggedright\sl#1\/}\else\relax\fi}
\reversemarginpar

\begin{document}
\title{High-Resolution Mid-Infrared Observations of High Mass Protostellar Objects}
 \author{James M. De Buizer}
\affil{Gemini Observatory, Casilla 603, La Serena, Chile}

\begin{abstract}
This article presents a review of all high-resolution ground-based mid-infrared observations of high mass protostellar objects (HMPOs) to date. An introduction to the characteristics of HMPOs and their relationship to the hot molecular core phase of massive star formation is outlined. I explain why observations in the mid-infrared of these types of sources are important, and in particular why high-resolution observations are needed. A summary is given of the earliest mid-infrared detections of HMPOs, followed by a detailed account of the most recent observations.   

\end{abstract}

\section{Introduction}

A {\it high mass protostellar object} (HMPO) is considered to be a protostar with a mass greater than about 8 solar masses in the earliest observable stage of formation. Young stars this massive are eventually capable of producing copious amounts of UV flux that will ionize the gas around them and form {\it ultracompact HII} (UC HII) regions. However, the HMPO phase precedes even this youthful UC HII phase of massive star formation. HMPOs are characterized by strong dust (i.e. sub-mm and mid-infrared) emission and molecular gas emission from the dense accreting envelope that surrounds and embeds the central massive protostar. Therefore, an HMPO will display molecular line emission at radio wavelengths, but lacks free-free emission from the ionized gas of a UC HII region. 

\begin{figure}
\plotone{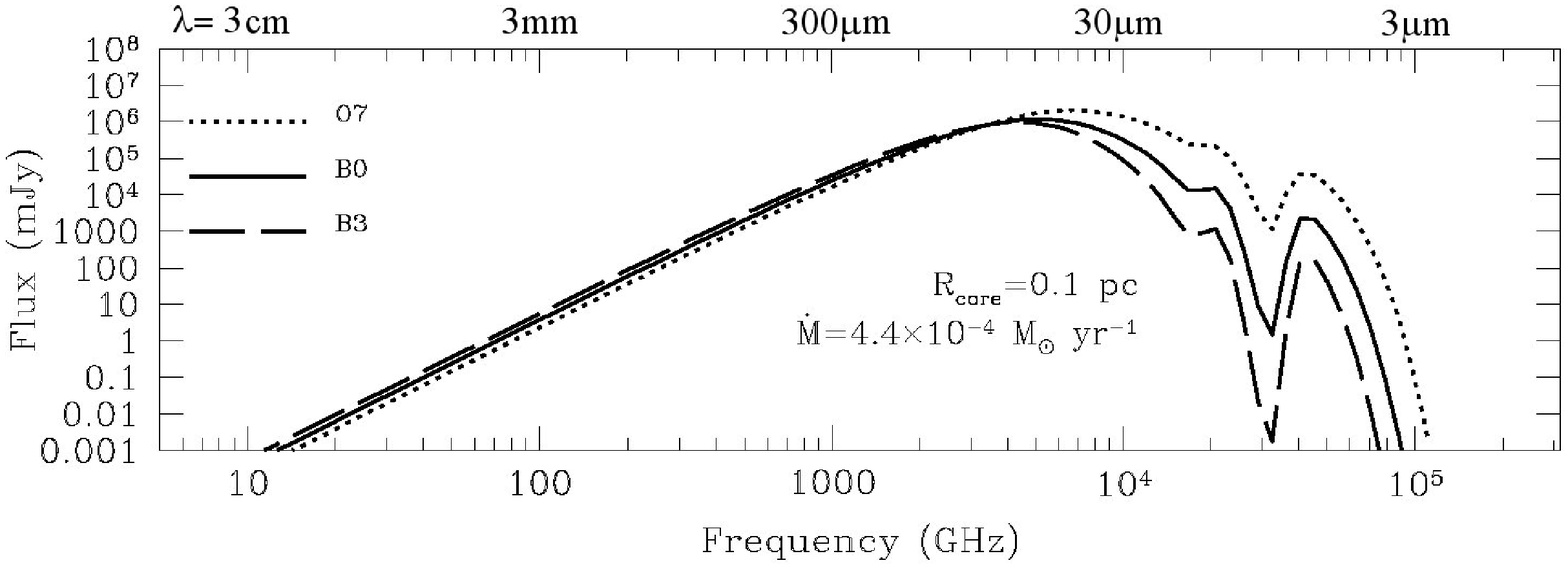}
\caption{Hot molecular core SED models from Osorio et al. (1999). This figure shows that when the luminosity is varied, the HMC SED changes significantly in the mid-infrared (3-30 $\mu$m), but not significantly at longer wavelengths. Varying accretion rate and core size also effect the shape and features of the SED in the mid-infrared (see examples in Osorio et al. 1999).}
\end{figure}

The HMPO definition encompasses several phases of massive protostellar formation described in the literature. It begins with a massive protostar just after creation via free-fall collapse, i.e. the {\it cold core} phase, which later evolves to transition to a {\it hot core} (a.k.a {\it hot molecular core}). By definition, both the HMPO and hot molecular core (HMC) phases end with the onset of a UC HII region. HMPOs younger than the HMC phase are very cold and deeply embedded in their envelopes and not detectable at mid-infrared wavelengths or shorter. Therefore, the only HMPOs that display mid-infrared emission are in the HMC phase of evolution. 

HMCs are observed to be compact sources ($\sim$0.1 pc). They are often found to be associated with maser emission and other high temperature and high density molecular tracers. They typically have internal kinetic temperatures greater than 150 K.  

HMCs began gaining attention and legitimacy as the precursors of massive stars after Cesaroni et al. (1994) performed high spatial resolution NH$_3$(4,4) imaging of four HMC sources found in the regions of G9.62+ 0.19, G10.47+0.03, G29.96-0.02, and G31.41+0.31. However, observations of some HMCs do exist before the work of Cesaroni et al. (1994). Most important to the discussion here is the work of Keto et al. (1992), who performed the first high spatial resolution mid-infrared imaging of two HMC sources, W3(OH) and G34.26+0.15, using one of the first mid-infrared cameras ever built.  

\section{Why observe HMCs in the mid-infrared?}

The mid-infrared emission ($\lambda$=5--25 $\mu$m) from HMCs can be a useful tool. Recently, the first models of the HMC phase of massive star formation have been constructed (i.e. Osorio et al. 1999). As can be seen in Figure 1, by changing parameters such as spectral type, accretion rates, and core radius, one can construct a variety of model spectral energy distributions (SEDs) for HMCs. These SEDs display little emission from wavelengths longer than 1 cm or shorter than 3 $\mu$m, just as seen in observations.  Throughout an array of physically reasonable parameter space, the SEDs peak around 80 $\mu$m on average. Notice in Figure 1 that on the Rayleigh-Jeans side of the SED ($>$80 $\mu$m), there is little change in the shape when spectral type is changed. However, on the Wien side of the SED ($<$80 $\mu$m), there is considerable change in SED shape. Most noticeable are the changes between 3 and 30 $\mu$m, and in the depth and shape of the feature at 10 $\mu$m. Varying accretion rate and core size also effect the shape and features of the SED in the mid-infrared.

This modelling shows that the mid-infrared is valuable in understanding the properties of the youngest massive stars. Only between the wavelengths of 3 and 30 $\mu$m can one constrain these types of massive star formation models. Therefore, if one can create an SED from observations of an HMC in the mid-infrared, one can use these models to back-out information and for the first time derive accurate physical parameters, such as mass, luminosity and accretion rate for a young and embedded massive star.

Unfortunately, there are not a lot of large-scale infrared surveys of these early phases of massive stellar birth. There have been three satellites capable of surveying massive star-forming regions in the mid-infrared: InfraRed Astronomical Satellite (IRAS), Infrared Space Observatory (ISO), and the Midcourse Space eXperiment (MSX). However the maximum spatial resolutions at 12 $\mu$m for these satellites were approximately 30$\arcsec$, 4$\arcsec$, and 18$\arcsec$, respectively. Therefore, the observations of the regions containing HMPOs that were made by these satellites were not particularly high-resolution.

\section{Why are high resolution observations of HMCs needed?}

Massive stars form in a highly clustered way.  Massive star-forming regions typically lie several kiloparsecs away, and therefore the apparent separation between the individual members is small. Because they evolve quickly, massive stars in several different stages of formation may be present in a massive star-forming region at one time. In fact, HMCs were determined to be precursors to massive stars partly due to their close proximity to UC HII regions. Low resolution observations at different wavelengths of the same massive star-forming region may very likely trace emission from different sources a different evolutionary states. Thus high-resolution observations are essential to resolve HMCs from other young nearby sources, and to be sure that observed emission at different wavelengths is in fact coming from the same source.

This resolution constraint limits the usefulness of mid-infrared satellite data in the study of individual HMCs. In practice therefore, we are restricted for now with ground-based mid-infrared observations of HMCs. Ground-based telescopes can yield diffraction-limited mid-infrared images on 4 to 10-m class telescopes, resulting in sub-arcsecond spatial resolutions. Therefore, in keeping with the theme of this symposium, I restrict the discussion in this article to ground-based mid-infrared observations.

\section{Early observations of HMCs}

As mentioned previously, the very first high-resolution mid-infrared observations of HMCs came from the work of Keto et al. (1992). They observed W3(OH), a well-studied region containing an HMC seen in HCN emission and an HII region (Turner \& Welsh 1984). This HMC has come to be known as W3(H$_2$O) or the {\it Turner-Welsh Object}. Keto et al. (1992) claimed to detect 12 $\mu$m emission from not only the HII region but the HMC as well. 

G34.26+0.15 was also observed by Keto et al. (1992) in both ammonia line imaging and at 12 $\mu$m. This region contains a well-studied cometary-shaped UC HII region. Keto et al. (1992) detected a core of ammonia a few arcseconds in front of the apex of the cometary UC HII region. Their 12 $\mu$m mid-infrared images revealed mid-infrared emission coming from the ammonia core location. 

Because the data in Keto et al. (1992) was taken with one of the first mid-infrared cameras, many years past until the next observations of HMCs were taken from the ground. This is most likely due to the fact that the mid-infrared array technology was quite new in 1992. It took a while for these arrays to be perfected, more readily available at astronomical observatories, and then reapplied to these types of objects. Therefore, the next publication of high resolution ground-based observations of HMCs was not until 2002. 

In the meantime, the study of HMPOs through their radio wavelength molecular line emission flourished in the decade of the 90s. These extensive observations revealed high abundances of high-excitation lines of several molecular species and established the name ``hot {\it molecular} core'' for these later phases of HMPO evolution. Important work during this period was made by Cesaroni et al. (1994, 1998), Olmi et al. (1996) and Gibb et al. (2000), among others, who made detailed studies of the chemical composition through molecular emission from these sources.  

\section{Recent HMC observations in the mid-infrared} 

The last few years have seen the beginning of a wealth of mid-infrared observations of HMCs. Wide-spread availability of mid-infrared cameras on the largest telescopes in the world have opened up the field and have lead to the recent flurry of high-resolution mid-infrared observations. 

We begin with the observations by De Buizer et al. (2002) of G29.96-0.02 -- the first publication of a mid-infrared detection of an HMC in the decade following the work of Keto et al. (1992). This region is home to a well-studied cometary-shaped UC HII region, and an HMC that was first imaged in ammonia line emission in the article of Cesaroni et al. (1994). De Buizer et al. (2002) performed sub-arcsecond resolution mid-infrared observations of the region of G29.96-0.02 on the Gemini 8-m telescope and detected the mid-infrared emission from the HMC, however it was barely resolved from the nearby UC HII region (Figure 2). These observations are a perfect example for the argument that high spatial resolution observations are indeed necessary to resolve individual sources in these massive star-forming regions.

Later in 2002, an article was published by Walsh discussing the results of previous multi-wavelength observations of G305.20+0.21. Using all the data available of the region to date, Walsh (2002) conjectures that the nature of the source found coincident with the methanol masers in this region is that of an HMPO. Firstly, there was no radio continuum emission found coming from the location of the methanol masers with an upper limit of 0.27 mJy at 4.8 GHz, though radio continuum was found from a UC HII region in this field (thus establishing it as a massive star-forming region). Secondly, observations at 10 an 20 $\mu$m yielded a detection of a compact source at the location of the methanol masers  (Figure 2). 

The mid-infrared emission from this maser source is very bright at both 10 and 20 $\mu$m. Using these two fluxes alone one can calculate a mid-infrared luminosity for this protostellar source (see De Buizer 2000 for method). One can then use this derived mid-infrared luminosity as an lower-limit approximation of the bolometric luminosity -- a gross underestimate, since HMPOs are known to have copious amounts of sub-mm flux. For G305.20+0.21, the lower-limit estimation of the bolometric luminosity is equivalent to that of a spectral type B1 star (De Buizer, Pina \& Telesco 2000). In other words, the exciting source of the methanol maser emission in G305.20+0.21 is at least as energetic as B1 star, and is quite possibly an O star (assuming there is only one exciting source at the center). A massive star this bright should be ionizing its surroundings and have a UC HII region, unless it is so young that the accreting envelope is quenching the expansion. Therefore, Walsh (2002) suggests that the exciting source of the methanol masers in G305.20+0.21 is an embedded B or O type star at a phase just before the onset of a UC HII region -- in other words, an HMC\footnote{Walsh (2002) also presents near-infrared observations that show emission at wavelengths as short as 1 $\mu$m apparently coincident with this HMC. This is unexpected due to the deeply embedded nature of HMCs. One possible explanation is that if there is an outflow from this source, the outflow can create a cavity in the HMC envelope which one can see direct or reflected near-infrared emission from inside, if the orientation is favorable.}. 
   
Also in 2002, Stecklum et al. published an article of their mid-infrared observations of W3(OH) from the Hale 5-m telescope. Stecklum et al. (2002) reported that they detected and resolved the HII region of W3(OH) itself, but they do not detect any mid-infrared emission from the HMC that was seen by Keto et al. (1992). Their upper limit on the detection of the HMC at 11.7 $\mu$m is well below the quoted integrated flux for this source of 45 mJy at 12.2 $\mu$m by Keto et al. (1992). Stecklum et al. (2002) did however find a second mid-infrared source, a UC HII region, offset northeast of W3(OH). Though this second mid-infrared source is $\sim$6$\arcsec$ northwest from the location of the HMC, Stecklum et al. (2002) conjectures that Keto et al. (1992) might have mistook this this second UC HII region for the HMC.  

In early 2003, Persi et al. published their infrared study of the region of G9.62+0.19 -- one of the four HMCs of Cesaroni et al. (1994). This region is rich in massive stellar sources in various stages of formation. Persi et al. (2003) claim to detect not only mid-infrared emission coming from the HMC, but near-infrared emission as well. However the method employed to register the HMC ammonia images with respect to the mid-infrared images was only accurate to a couple of arcseconds.

The most recent paper on HMCs as of the writing of this article is that of De Buizer et al. (2003). They performed high-resolution mid-infrared imaging towards seven HMCs and HMC candidates using the IRTF 3-m telescope ($\sim$1$\arcsec$ resolution), the Gemini North 8-m telescope ($\sim$5$\arcsec$ resolution), and the Keck II 10-m telescope ($\sim$0.4$\arcsec$ resolution). There are several sources in common between this small survey paper and the work by others already discussed. 

For instance, De Buizer et al. (2003) observed G9.62+0.19 and were able to detect at least four sources on the field that are also seen in cm radio continuum emission. Because there is hardly any astrometrical error between the ammonia line images and the radio continuum images, having four sources in common on this field in the mid-infrared and radio allows for the first time extremely accurate registration of the mid-infrared with the ammonia emission in this region. Given this accurate relative astrometry, it was found that the mid-infrared emission thought to be coming from the HMC is in fact not, but is instead coming from a separate source 2$\arcsec$ away. Since HMCs are so embedded that one would not expect any near-infrared emission from them, this mid-infrared source is also most likely the origin of the near-infrared emission seen by Persi et al. (2003).

\begin{figure}
\plotfiddle{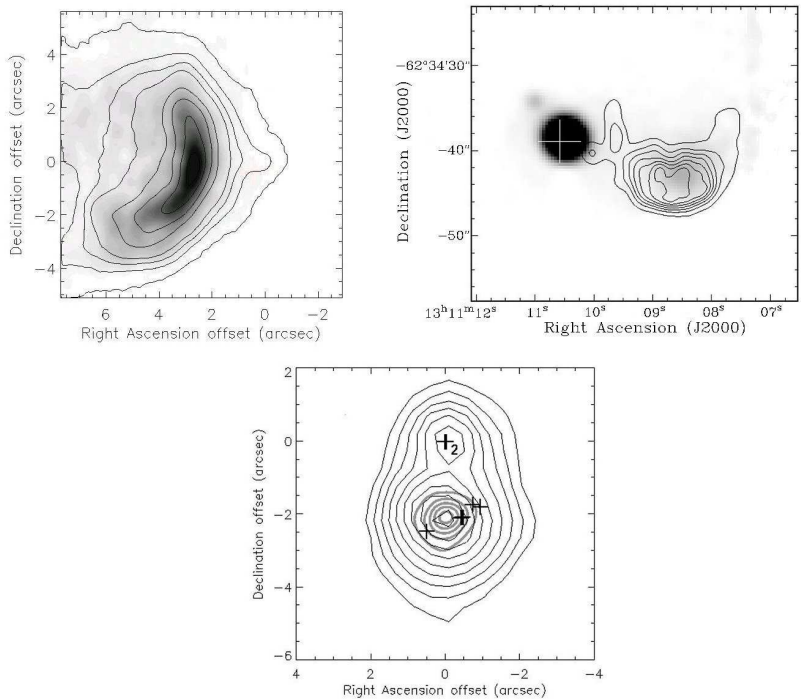}{8.2cm}{0}{48}{48}{-170}{-23}
\caption{Three mid-infrared-bright HMCs. {\bf(Upper Left):} G29.96-0.02 (De Buizer et al. 2002). The HMC can be seen as a bump in the 18 $\mu$m emission (contours) of the UC HII region. Gray-scale shows the 24 GHz continuum observations of Cesaroni et al. (1998) which delineate the cometary UC HII region only. {\bf(Upper Right):} G305.20+0.21 (Walsh, priv. comm.). The HMC is a luminous 10 $\mu$m source (gray-scale) coincident with the maser location (cross).  Contours show 4.8 GHz continuum emission nearby but not coincident with this HMC. {\bf(Bottom):} G45.07+0.13 (De Buizer et al. 2003). The HMC is the northern of the two luminous 11.7 $\mu$m (thin black contours) sources and coincident with the masers (crosses) marked with a 2. The southern source is a UC HII region, evidenced by the the 15 GHz continuum (thick gray contours; Hofner \& Churchwell 1996).}
\end{figure}

I already discussed the Stecklum et al. (2002) result where they did not detect the HMC that Keto et al. (1992) detected near W3(OH). The second source from Keto et al. (1992), G34.26+0.25, was also among the sources observed by De Buizer et al. (2003). Like the HMC in W3(OH), this 70 mJy HMC seen at 12.5 $\mu$m by Keto et al. (1992) was not detected in the observations of De Buizer et al. (2003) with 3-sigma a upper limit flux density of 4 mJy through a broadband 10 $\mu$m (N) filter. 

Two new HMC candidates were found in the survey by De Buizer et al. (2003). G45.07+0.13 is a field that contains a compact UC HII region that has water maser emission. However, offset 2$\arcsec$ north of the UC HII region is an isolated group of water masers with no associated radio continuum emission (Hofner \& Churchwell 1996). De Buizer et al. (2003) detected a extremely bright mid-infrared source at this maser location at both 11.7 and 20.8 $\mu$m (Figure 2). From the mid-infrared luminosity alone, they derived a lower limit  bolometric luminosity that is equivalent to a spectral type B0 star. This means that this source in G45.07+0.13 is an (even more massive) analog to the high mass protostellar object seen in the G305.20+0.21 region. 

The second HMC candidate was found in the region of G11.94-0.62. Like G45.07+0.13 and G305.20+0.21, this region was imaged in the mid-infrared because it contained a UC HII region with a group of isolated water masers offset. The mid-infrared observations of De Buizer et al. (2003) revealed that there is indeed a mid-infrared source at this maser location. The mid-infrared luminosity yields a lower limit to the bolometric luminosity of B9 for this source. However, extinction may play a role in masking the true mid-infrared flux of this source, and it likely houses a much more massive protostar. Further evidence that this source is an HMC comes from the spectroscopic observations that show high-excitation molecular line emission coming from this region, a necessary (but not sufficient) condition if this is indeed an HMC. However, further observations of the HMC candidate in G11.94-0.62 are needed to ascertain its true nature.

\section{Evolutionary status of mid-infrared-bright HMPOs/HMCs}

Of the eight regions containing HMCs or HMC candidates observed by De Buizer et al. (2002) and De Buizer et al. (2003), mid-infrared emission was found coming from only three. Though it is small number statistics, this suggests a success rate of 38\% when searching for mid-infrared emission from HMCs. It is likely that the remaining undetected sources are still so embedded that they cannot yet be seen via their mid-infrared emission. This implies that mid-infrared-bright HMCs are more evolved than mid-infrared-dark HMCs. If this is the case, we may be able to use the mid-infrared to determine if an HMC source is at a stage just before onset of a UC HII region.

However, it is also possible that the extinction towards these regions is generally so high as to decrease the probability of detection of the HMCs in the mid-infrared as well. Spectroscopic observations towards these sources, for example, would help to determine the amount of extinction present. 

\section{Conclusions}

Observations of HMPOs are extremely important tools in understanding the birth and early evolution of O and B type stars. Though observations at all observable wavelengths are valuable, the mid-infrared allows a unique opportunity to observe the warm dust emission of these embedded massive protostars. Also, construction of the mid-infrared portion of the SEDs for such sources may prove invaluable in the testing of massive star formation models and learning about the physical conditions of massive embedded protostars.

High-resolution ground-based observations are almost a necessity for mid-infrared HMPO/HMC study. For instance, in the cases of G29.96-0.02 and G45.07+0.13, observations required a 8-m and 3-m telescope, respectively, in order to just resolve the HMCs from the mid-infrared emission of nearby UC HII regions. With no mid-infrared satellite telescopes of this size available in the near future, the detailed study of HMPOs and HMCs will have to continue to be conducted from the ground.

There are apparently only a few bona-fide mid-infrared-bright HMCs known at present: G29.96-0.02, G45.07+0.13, and G305.20+0.21. Contrary to previous beliefs, mid-infrared emission cannot be confirmed to be associated with the HMCs in G9.62+0.19, G34.26+0.15, nor W3(OH). 

Though more sources need to be studied, it appears that HMCs with mid-infrared emission are most likely a more evolved phase of HMC close to the beginning of the UC HII phase. Because there are so few mid-infrared bright HMCs known, not much more can be said for the global properties of the population. However, if the rate of observations of these sources over the last three years is an indication of things to come, we shall be learning much more about the nature and physical characteristics of HMCs in the very near future.

\end{document}